\documentclass[11pt,a4paper]{article} 
\pdfoutput=1 %
\usepackage{jcappub} %

\usepackage{amssymb}
\usepackage{graphics}
\usepackage{epsfig}
\usepackage{graphicx}
\usepackage{subfig}
\usepackage{bm}

\graphicspath{ {./figures/}{./figures/sliceplot/}{./figures/sliceplot/slices-nonabelian/}{./figures/sliceplot/slices-abelian/}{./figures/backreactiontime/}{./}{./figures/nk/}{./figures/energies/}{./figures/plaquette/}{./figures/potcomp/}{./figures/nknonabelian/} }

\newcommand{\ud}{\mathrm{d}}

\newcommand{\mbf}[1]{\mathbf{#1}}

\newcommand{\chiosc}{\chi_{\mathrm{osc}}}

\def\beq{\begin{equation}}
\def\eeq{\end{equation}}
\def\baq{\begin{eqnarray}}
\def\eaq{\end{eqnarray}}

\title{Lattice Calculation of the Decay of Primordial Higgs Condensate}

\author[a]{Kari Enqvist,} 
\author[a,b]{Sami Nurmi,} %
\author[a]{Stanislav Rusak,} %
\author[c]{and David J. Weir} %
\affiliation[a]{University of Helsinki and Helsinki Institute of Physics, P.O. Box %
64, FI-00014, Helsinki, Finland } %
\affiliation[b]{Department of Physics, University of Jyv\"{a}skyl\"{a}, P.O.
Box 35 (YFL), FI-40014 University of Jyv\"{a}skyl\"{a}, Finland}
\affiliation[c]{Faculty of Science and Technology, University of Stavanger, 4036 Stavanger, Norway} %
\emailAdd{kari.enqvist@helsinki.fi} %
\emailAdd{sami.nurmi@helsinki.fi} %
\emailAdd{stanislav.rusak@helsinki.fi} %
\emailAdd{david.weir@uis.no} %
\keywords{} %
\arxivnumber{} %

\abstract{We study the resonant decay of the primordial Standard
Model Higgs condensate after inflation into $\mathrm{SU}(2)$ gauge
bosons on the lattice. We find that the non-Abelian interactions
between the gauge bosons quickly extend the momentum distribution
towards high values, efficiently destroying the condensate after the
onset of backreaction. For the inflationary scale $H = 10^{8}$ GeV,
we find that $90$\% of the Higgs condensate has decayed after $n\sim
10$ oscillation cycles. This differs significantly from the Abelian
case where, given the same coupling strengths, most of the
condensate would persist after the resonance.}

\preprint{HIP-2015-20/TH}

\begin{document}

\maketitle 

\section{Introduction}

After the detection of Standard Model (SM) Higgs
\cite{ATLAS:2012ae,Chatrchyan:2012tx} there has been a growing
interest in the cosmological role of the Higgs field $\Phi$. Intriguingly, the Higgs might even
act as the inflaton if its potential at high energies is dominated by  a coupling to the spacetime curvature
of the form $\xi\Phi^{\dag}\Phi R$ \cite{Bezrukov:2007ep}. The connection to the low energy SM regime
of the scenario is, however, not unique due to its non-renormalizability \cite{Bezrukov:2014ipa,Bezrukov:2014bra,Hamada:2014iga} and the required large coupling
$\xi\gtrsim 10^{3}$ could be subject to questions of naturalness \cite{Barbon:2009ya,Burgess:2010zq,
Bezrukov:2010jz}.

However, the Higgs could also be of cosmological interest in several
other ways. If the Higgs potential remains close to the SM form
up to the inflationary scale  $H_*$,  during inflation the Higgs would be a light spectator
field, as discussed in
\cite{Espinosa:2007qp,DeSimone:2012qr,Enqvist:2013kaa,Enqvist:2014bua,Kobakhidze:2013tn,Fairbairn:2014zia,Hook:2014uia,Espinosa:2015qea}. In such a case inflation needs
to be driven by some new physics beyond SM. Like for all light fields during
inflation, both the Higgs mean field as well as the local field
value will be subject to fluctuations. The latter are isocurvature
perturbations during inflation while fluctuations of the mean field
generate an effective Higgs condensate. The condensate survives
inflation and sets specific non-vacuum initial conditions for the
hot Big Bang epoch \cite{Espinosa:2007qp,Enqvist:2013kaa}. The typical magnitude of
the Higgs mean field is, provided inflation lasts long enough, of
the order of $H_*$. Depending on details of
physics beyond SM the primordial Higgs condensate could have
significant observational impacts ranging from imprints in CMB to
baryogenesis and dark matter \cite{DeSimone:2012qr, DeSimone:2012gq,Choi:2012cp,Enqvist:2014zqa, Kusenko:2014lra}.

Theoretical self-consistency of the setup requires a stable vacuum. As is well
known, the pure SM vacuum in flat space becomes unstable above
energies $\mu_{\rm c}\sim 10^{10}$ GeV for the measured best fit
values of the strong coupling constant, the top mass, and the Higgs
mass $M_{h}\simeq125$ GeV
\cite{Espinosa:2007qp,Bezrukov:2012sa,Degrassi:2012ry,Buttazzo:2013uya, Kobakhidze:2013tn, Fairbairn:2014zia,Hook:2014uia,  Kobakhidze:2014xda,Spencer-Smith:2014woa, Espinosa:2015qea}.
However, as discussed in \cite{Espinosa:2007qp,Herranen:2014cua}, during inflation the
stability depends crucially also on the Higgs-curvature
coupling $\xi\Phi^{\dag}\Phi R$. This coupling is necessarily
generated by radiative corrections even if $\xi$ would be set to
zero at some scale. A one-loop investigation \cite{Herranen:2014cua}
shows that vacuum stability can be maintained even within the SM up
to the maximal  inflationary scale $H_*\lesssim
10^{14}$ GeV consistent with the tensor bound \cite{Ade:2014xna,Ade:2015xua},
provided that $\xi_{\rm EW} \gtrsim 0.1$ at the electroweak scale.

Thus quite generically, immediately after inflation the Universe
features a primordial Higgs condensate. In the case of Higgs
inflation, in which the Higgs itself sets the dynamics, the Universe
is dominated by the Higgs condensate, and its decay time directly
determines the reheating temperature \cite{Bezrukov:2008ut,GarciaBellido:2008ab}. On the other hand, within the SM, and in a broad class of its
extensions, the condensate contributes very little to the energy
density of the Universe. At the onset of hot Big Bang its large
displacement from the SM vacuum could nevertheless have important
ramifications for the subsequent evolution of the Universe  \cite{DeSimone:2012qr,DeSimone:2012gq,Enqvist:2014zqa,Kusenko:2014lra}. Thus even if the Higgs were a mere spectator during inflation, it can
affect post-inflationary physics until the condensate eventually
decays. In both cases a detailed understanding of the Higgs
condensate decay is crucial in order to properly address the observational
imprints.

At zero temperature the dominant decay channel for both the SM Higgs
and the non-minimally coupled Higgs inflaton is the non-perturbative
production of gauge bosons through a parametric resonance
\cite{Bezrukov:2008ut,GarciaBellido:2008ab,Enqvist:2013kaa,Enqvist:2014tta,Figueroa:2015rqa}. The
investigation of the resonance is complicated by the non-Abelian
dynamics of the gauge field which have not been carefully explored
so far. In the regime of broad resonance, realized for SM running of
the gauge couplings, the non-Abelian terms are small during the
first stages of the resonance but rapidly grow important as the
exponential production of gauge fields starts to backreact on the
dynamics of the Higgs condensate. The efficiency and duration of the
resonant Higgs decay therefore cannot be reliably estimated without
properly accounting for the non-Abelian couplings. In the limit of a
narrow resonance, which could occur in specific extensions of the
SM, the non-Abelian features affect the resonance dynamics already
before the dynamical backreaction of gauge fields
\cite{Enqvist:2014tta}.

In this work we investigate the resonant decay of the
Higgs into gauge bosons using numerical lattice simulations and accounting for the full non-Abelian dynamics.
In particular, we compute the decay time of the subdominant SM Higgs
condensate generated during inflation.

The paper is organized as follows. In Section \ref{sec:SM} we
present the framework to investigate the resonant Higgs decay and
make some analytical estimates. In Section \ref{sec:lattice} we
present the results of our full lattice computation of the Higgs
decay which constitutes the main part of the paper. Finally we
conclude in Section \ref{sec:conclusions}.

\section{Resonant Higgs decay in Standard Model}
\label{sec:SM}

During and shortly after inflation the energies are significantly
higher than the electroweak scale. Thus the Standard Model Higgs
potential can be approximated by the quartic term alone. The relevant part
of the SM action for our discussion is
\begin{equation}
S = -\int \ud^4x \left\{\frac{1}{4} \eta^{\mu\alpha}\eta^{\nu\beta}\Big(F_{\mu\nu}^a F_{\alpha\beta}^a + G_{\mu\nu}G_{\alpha\beta}\Big) +
a^4\left[\left(D_{\mu}\Phi\right)^{\dagger}D^{\mu}\Phi + \lambda(\Phi^{\dag}\Phi)(\Phi^{\dag}\Phi)^2\right]
\right\}\
\end{equation}
with the kinetic terms
\begin{eqnarray}
F^a_{\mu\nu} &=& \nabla_{\mu}A_{\nu}^a - \nabla_{\nu}A_{\mu}^a +
g\epsilon^{abc}A_{\mu}^bA_{\nu}^c\ , \\\nonumber
G_{\mu\nu} &=&
\nabla_{\mu}B_{\nu} - \nabla_{\nu}B_{\mu}\ ,\\\nonumber
D_{\mu}\Phi &=& \left(\nabla_{\mu} -igA^a_{\mu}\tau^a -
\frac{i}{2}g'B_{\mu}\right)\Phi\ .
\end{eqnarray}
Here $\Phi$ is the Higgs doublet and $A_{\mu}^a$ and $B_{\mu}$ are
the $\mathrm{SU}(2)$ and $\mathrm{U}(1)$ gauge fields, respectively.

We assume that the inflationary stage is driven by some physics
beyond the SM. To reheat the universe, the inflaton field, or
fields, should arguably be coupled to the SM degrees of freedom
through non-gravitational interactions. We assume these couplings
are small and use the pure SM running for the Higgs and gauge
couplings. More importantly, we will be investigating the decay of
the Higgs field assuming there is no thermal background. If the
inflaton has already decayed at this stage, we are therefore
implicitly assuming that the interaction rate between its decay
products and the SM fields is small compared to the Hubble time over
our epoch of interest. On the other hand, one could argue that the
decay rate of the inflaton field would typically be much smaller
than the decay rate of the Higgs condensate. In such a case the
oscillating inflaton field would give rise to an effectively matter
dominated universe.

\subsection{Initial conditions from inflation}

Given the SM potential, the SM Higgs is both a light field and
energetically subdominant during inflation \cite{DeSimone:2012qr,Enqvist:2013kaa}.
Its super-horizon dynamics can be investigated using the stochastic
formalism \cite{Starobinsky:1986,Starobinsky:1994bd}. Starting from
a generic field configuration at some point during inflation, the
Higgs distribution relaxes to the equilibrium given by $P(h)\sim
{\rm exp}({-8\pi^2V/3H^4})$ within a time scale of $N_{\rm
rel.}\simeq 100$ e-folds \cite{Enqvist:2012xn}. Provided that inflation lasted  somewhat
longer we can therefore quite generally assume that the value of the
Higgs condensate in the observable universe is to be drawn from the
equilibrium distribution.

After the end of inflation the typical mean Higgs field value over the
observable universe is thus given by
  \beq
  \label{eq:equilibrium}
  h_{*}=\sqrt{\langle h^2\rangle} = 0.36 \lambda_{*}^{-1/4} H_{*}\ .
  \eeq
The inflationary stage thus generates a primordial Higgs condensate
$h_*$, setting specific non-equilibrium initial conditions for the
Hot Big Bang epoch. Here we are implicitly assuming a low
inflationary scale $H_{*}\lesssim 10^{9}$ GeV such that the Higgs
coupling to curvature scalar $\xi\Phi^{\dag}\Phi R$ can be neglected
\cite{Herranen:2014cua}.

\subsection{Onset of the resonant Higgs decay}

The Higgs condensate becomes effectively massive as $H^2_{\rm osc} =
3\lambda_{*} h_{*}^2$ and it starts to oscillate around the vacuum
expectation value $h=0$. In terms of the rescaled field $\chi=ah$
and conformal time $d\tau = a^{-1}dt$ the equation of motion reads
  \beq
  \label{eq:chiEOM}
  \ddot{\chi}+\lambda\chi^3-\frac{\ddot{a}}{a}\chi=0\ .
  \eeq
The last term can in general be neglected. In a radiation
dominated background it is identically zero and for matter
domination it becomes negligible soon after the onset of Higgs
oscillations. The expansion of space can therefore be scaled out
from the dynamics and the problem essentially reduces to flat space.

The resonant production resulting from coherent oscillations of
scalar fields has been extensively studied in the literature
\cite{Traschen:1990sw,Kofman:1994rk,Kofman:1997yn,Greene:1997fu}. Neglecting self-interactions, the weak transverse
components of the gauge fields obey the Lam\'{e} differential
equation
\begin{equation}
\label{eq:lame}
\frac{\ud^2\mbf A^T}{\ud z^2} + \left[\kappa^2 + q\operatorname{cn}^2\left(z,\frac{1}{\sqrt{2}}\right)\right]\mbf A^T = 0.
\end{equation}
where
\begin{equation}\label{newpara}
z \equiv \sqrt{\lambda \chiosc^2}(\tau-\tau_{\mathrm{osc}}), \qquad \kappa^2 \equiv \frac{k^2}{\lambda\chiosc^2},
\qquad q \equiv \left\{\begin{array}{cl}\frac{g^2}{4\lambda} & \text{for the W bosons} \\ \frac{g^2+g'^2}{4\lambda} & \text{for the Z boson}\end{array}\right.,
\end{equation}
and $q$ is the resonance parameter characterizing the strength of particle production. For the pure Standard Model
this parameter is always greater than unity, which puts the system in the broad resonance regime \cite{Enqvist:2014tta}. The Lam\'{e} equation exhibits instability regions where the solutions are amplified exponentially resulting
in explosive production of gauge boson quanta.

The exponential production of gauge fields could in principle be
washed out by rapid perturbative decays of the generated gauge
fields. This is indeed what happens in the early stages of reheating
after the Higgs inflation \cite{Bezrukov:2008ut,GarciaBellido:2008ab}. For
the energetically subdominant SM Higgs this is however not the case.
The perturbative decay rate of $W$ bosons into any pair of fermions
is small compared to the characteristic time scale of Higgs
oscillations
  \beq
  \frac{\Gamma_{W}}{m_{\chi}} =
  \sqrt{3}{32\pi}\frac{g^3}{\sqrt{\lambda}} = {\cal O}(0.01)\ .
  \eeq
Similar estimates hold for the $Z$ bosons. The perturbative decays
are therefore irrelevant compared to the resonant production of
gauge fields and can be neglected in what follows.

\subsection{Backreaction and limitations of the analytical approach}

In the discussion of the previous section we have ignored the
non-Abelian interactions of the gauge fields as well as the
backreaction of the produced particles on the dynamics of the Higgs
field. This is justified in the beginning of the resonance where the
gauge field occupation numbers are small.

As the occupation numbers grow larger than one, the non-Abelian
terms backreaction effects eventually start to dominate the
dynamics. Using the Hartree approximation one can estimate the onset
of backreaction by evaluating the time when \cite{Enqvist:2014tta}
  \beq
  \label{analytical_backreaction}
   \frac{g^2}{2}\langle
   W^{+}_{\mu}W^{\mu-}\rangle+\frac{g^2+g'{}^{2}}{4}\,
  \langle Z_{\mu}Z^{\mu}\rangle \sim 3 \lambda \chi^2\ .
  \eeq
The backreaction destroys the coherence of the Higgs oscillations
and eventually shuts down the resonance
\cite{Kofman:1994rk,Kofman:1997yn,Greene:1997fu}. The dynamics in
this regime is entirely dominated by interactions and its analysis
requires a full non-linear lattice computation. Before turning to
the numerical results in the next section let us however present
some qualitative comments on the role of non-Abelian interactions.

For Abelian fields the distribution of occupation numbers would
retain its form peaked around $\kappa\sim q^{1/4}/2$ and essentially a
vacuum configuration for $\kappa\gg q^{1/4}$. The low energetic gauge
bosons and Higgs particles $\kappa\lesssim q^{1/4}$ would be inefficient
in fragmenting the remnants of the Higgs condensate. Consequently,
it would seem that a significant part of the condensate would
survive long after the backreaction.

The non-Abelian interactions of the $W$ bosons however completely
change the picture. As the occupation numbers around the resonant
peak $\kappa\sim q^{1/4}/2$ grow much larger than unity the $W$
scatterings and annihilations into $Z$ bosons rapidly generate
excitations with higher momenta. This generates a bath of gauge
bosons extending to $\kappa\gg q^{1/4}$ which would appear to efficiently destroy the
remnants of the Higgs condensate, in sharp contrast with the Abelian case.
As we will discuss below, this behaviour is exactly what we see in our
numerical computation.

We should note here that also annihilations of the gauge bosons into
two fermions participate in the Higgs decay. Indeed, the vacuum
cross sections of the processes $WW\rightarrow f{\bar f}$ and
$WW\rightarrow WW$ both scale as $\sigma \sim \alpha_{W}^2/m_{W}^2$
and the production of light fermions is also kinematically favoured,
as discussed in the context of preheating after the Higgs inflation
\cite{Bezrukov:2008ut,GarciaBellido:2008ab}. However, at the final stages
of the resonance the non-Abelian interactions $WW\rightarrow WW,
WW\rightarrow Z$ are significantly favoured by the large occupation
numbers $n_{W,Z}\>> 1$. In our lattice computation we will therefore
neglect the decay channels into light fermions.

\section{Full lattice computation of the Higgs decay}

\label{sec:lattice}

Motivated by the considerations above, we simulate the SM Higgs
dynamics on lattice during the period over which exponential
production of gauge bosons is taking place. As the Higgs decay
proceed dominantly through the production of non-Abelian $W$ bosons
we simplify the problem by investigating a $\mathrm{SU}(2)$
gauge-Higgs system instead of the full $\mathrm{SU}(2)\times
\mathrm{U}(1)$ symmetry. Technical details of the lattice simulation
can be found in Appendix~\ref{sec:latticedetails}. The techniques we
use are similar to those employed in previous numerical studies of
preheating involving gauge fields~\cite{Rajantie:2000nj,DiazGil:2008tf,Skullerud:2003ki}.

Note that the system under consideration is conformal and so the
effect of expansion can be neglected. Therefore there is no mass
scale, other than that set by the initial value of the Higgs field.

Our initial conditions for the
gauge field are Gaussian white noise with strength $\xi$,
\begin{equation}
  \langle A_i^a(k,0) A_j^b(k',0) \rangle = \xi \delta^{ab} \delta_{ij}
  \delta(k-k')\ .
\end{equation}
rather than the quantum vacuum. To avoid needing to carry out
projection to satisfy the Gauss law initially, we assume that the
gauge field conjugate momentum is initially zero. The gauge field
equilibrates quickly, in any case, and loses all memory of these
initial conditions as the particle number starts to grow. We then
take as $z=0$ the time when the particle occupation number at the
peak of the resonance exceeds $n_\kappa = 0.1$.

We have confirmed that the normalization $\xi$ of the gauge
field initial conditions only has a logarithmic impact on the time
taken for the system to equilibrate. For the purposes of the
current paper, then, the above Gaussian noise initial conditions are adequate.

The lattice simulations are carried out in temporal gauge, which is
equivalent to setting $A_0 = 0$, however we must further fix lattice
Coulomb gauge to recover particle numbers. When we wish to measure
particle numbers in the gauge field we use standard gauge fixing
techniques (see Appendix~\ref{sec:gaufix}) and then measure the
connected two point
functions~\cite{Aarts:1999zn,Salle:2000hd,Skullerud:2003ki}
\begin{align}
\left< A_i^a (\mathbf{k},t) A_j^b(-\mathbf{k},t) \right> & =
\delta_{ab} \left[ \left( \delta_{ij} - \frac{k_i k_j}{k^2} \right) D_T^A(k,t)
  + \frac{k_i k_j}{k^2} D_L^A(k, t) \right] \label{eq:Aexpectation}, \\
\left< E_i^a (\mathbf{k},t) E_j^b(-\mathbf{k},t) \right> & =
\delta_{ab} \left[ \left( \delta_{ij} - \frac{k_i k_j}{k^2} \right) D_T^E(k,t)
  + \frac{k_i k_j}{k^2} D_L^E(k, t) \right]. \label{eq:Eexpectation}
\end{align}
With $D_L^A(k,t) = 0$ in Coulomb gauge we define the gauge field
particle number as
\begin{equation}
\label{eq:particlenumber}
n_k^A(t) = \sqrt{D_T^A(k,t)D_T^E(k,t)}.
\end{equation}
This quantity is not gauge invariant and so we also present the power
spectrum of the energy density in the gauge field in our results. This
observable does not show the clear sharp resonant peak but it does
allow one to identify backreaction and the onset of equilibration.

For comparison we also plot the results of a `non-compact'
$\mathrm{U}(1)$ simulation with equivalent initial
conditions\footnote{Note that a lattice simulation of `compact'
  $\mathrm{U}(1)$ would not correctly show the onset of backreaction
  as the photon field $A_i(x)$ would be symmetric under
  transformations $A_i(x) \to A_i(x) + 2n \pi$. For the same reason,
  it is not sufficient to only initialise one colour of the
  $\mathrm{SU}(2)$ gauge field to get an `abelian' comparison to the
  nonabelian case.}. A comprehensive numerical study of the resonant
production of Abelian gauge bosons was recently carried out in
Ref.~\cite{Figueroa:2015rqa}, and so we will not go into detail here.

In general we use lattices of volume $128^3$. With gauge fixing, each
simulation at this volume typically required four hundred CPU-hours;
in total the results of this paper required less than ten thousand
CPU-hours. On the infrared side we varied the lattice volume to ensure
that the parametric resonance was not unacceptably cut off at long
wavelengths, and confirmed that the results of this paper do not
depend on this. It is less straightforward to establish that the
physics does not depend on rescaling the system, because it is
conformal and there is no mass scale.

To provide a numerical estimate of the time taken for the Higgs field
to backreact, we define the \textsl{backreaction time} $z_\text{br}$
as the time when the amplitude of the Higgs oscillations falls to 90\%
of its initial value $h_*$. We emphasise, however, that this time is
logarithmically dependent on the amplitude $\xi$ of the initial
conditions for the gauge field. The simulations presented here are
however more than sufficient to demonstrate the correct qualitative
behaviour.

After the backreaction on the Higgs commences, the system starts to
equilibrate and the equation of state for the scalar field
approaches the equipartition value ($1/3$). We also see indications of
thermalisation. However, because we do not have a mass scale in our
simulations, we cannot give a definitive estimation of the
thermalisation time. This must be left to future work that fully
incorporates expansion.

\subsection{Results}

To give a broad resonance peak within our lattice volume we rescaled
the system so that the lattice spacing is unity and work with
rescaled lattice Higgs field $\Phi = 2\chi$.

\begin{figure}
\centering
\includegraphics[width=0.5\textwidth]{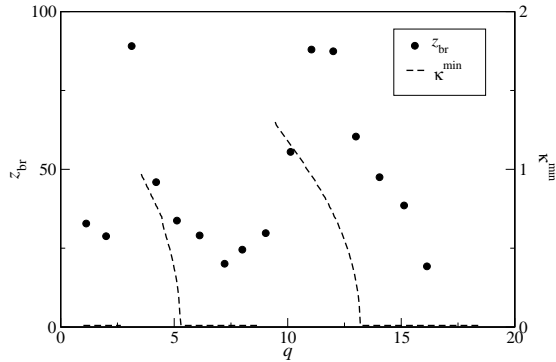}
\caption{\label{fig:backreactiontime}\small Plot of the backreaction
  time $z_\text{br}$ against $q$ for the non-Abelian theory. It seems
  that some fine tuning is required to substantially change the Higgs
  decay time. Also shown is the smallest $\kappa$ that lies within a
  resonance band at a given $q$ (for a full contour plot of resonance
  bands see e.g. Ref.~\cite{Enqvist:2014tta}).}
\end{figure}

In Figure~\ref{fig:backreactiontime}
we show how the backreaction time varies with $q$ for the noise
amplitude. For the Standard Model changes of $q=g^2/(4\lambda)$
correspond to changes of the inflationary scale $H_{*}$ which
determines the values of the running coupling constants, identifying
the renormalization scale as $\mu = h_* \sim H_*$; see
Eq.~(\ref{eq:equilibrium}). For example, taking $H_* = 10^{8}$ GeV
yields with the SM running and best fit inputs the coupling values
$g\simeq 0.6$ and $\lambda=0.15$ at the end of
inflation~\cite{Degrassi:2012ry}. This gives the resonance parameter
$q =g^2/(4\lambda)\sim 6$. The peaks (slowest to backreact) are
associated with the values of $q$ where the first (lowest-momentum)
resonant band encountered is farthest from $\kappa = 0$.

\begin{figure}
\centering
\vspace{-20pt}
\hspace{-50pt}\includegraphics[width=0.75\textwidth]{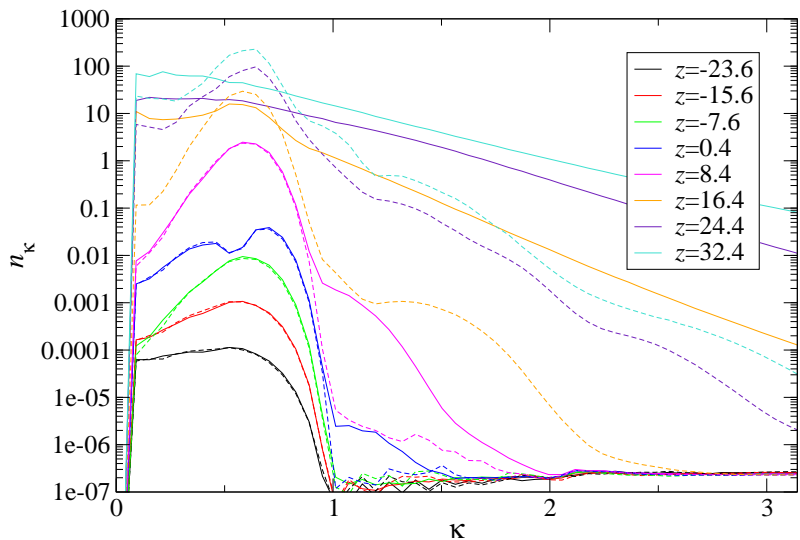}
\caption{\label{fig:comparison}\small Comparison of particle numbers
  $n_\kappa$ for abelian (dashed lines) and non-Abelian (solid lines)
  simulations ($q\approx 6.12$) evaluated at evenly-spaced
  time intervals. Equilibration in the abelian case is much slower and
  particle numbers remain many orders of magnitude smaller.}
\end{figure}

\begin{figure}
\centering
Nonabelian \\
\subfloat{\includegraphics[width=0.1\textwidth]{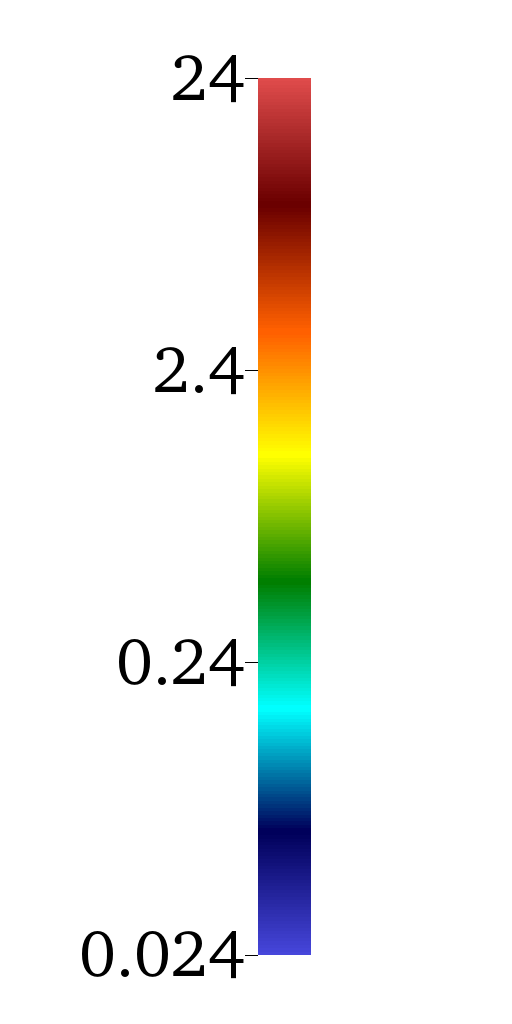}}
\subfloat{\includegraphics[width=0.2\textwidth]{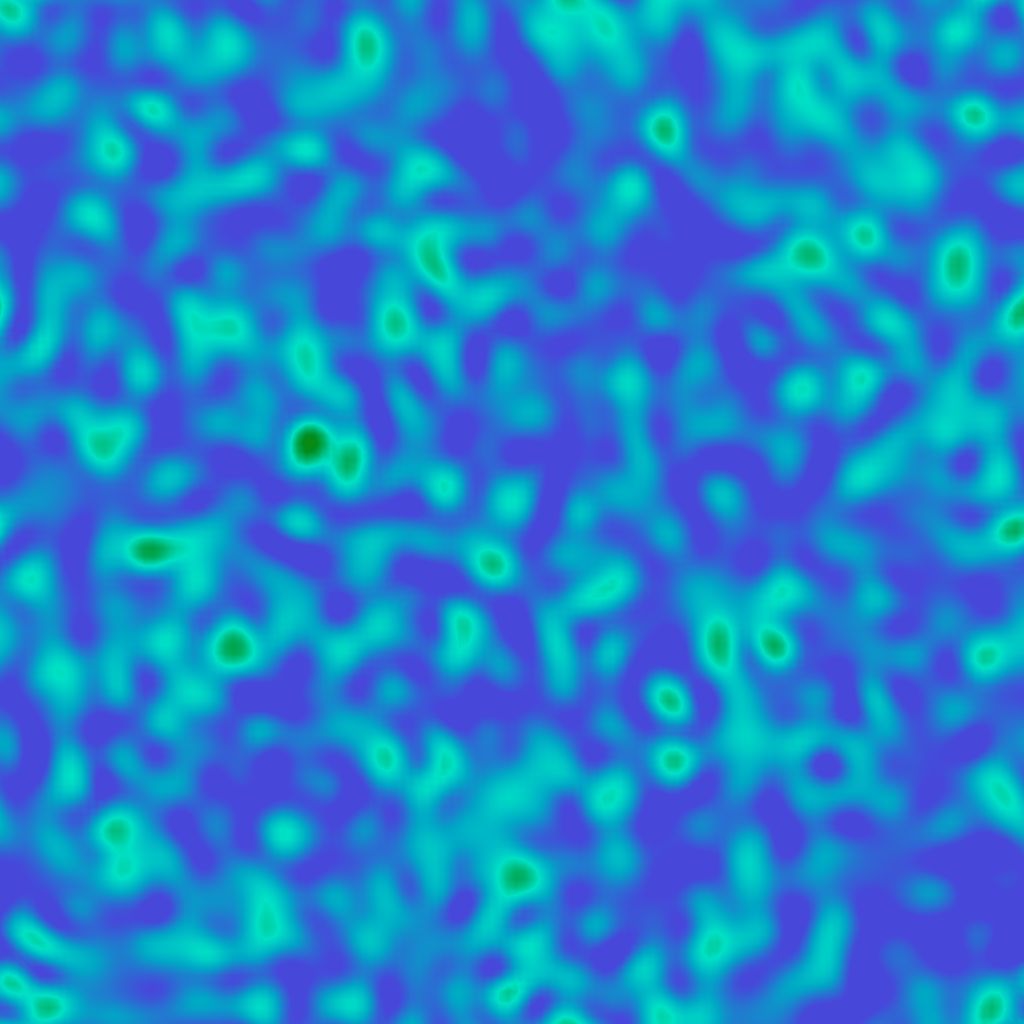}}
\hspace{5pt}
\subfloat{\includegraphics[width=0.2\textwidth]{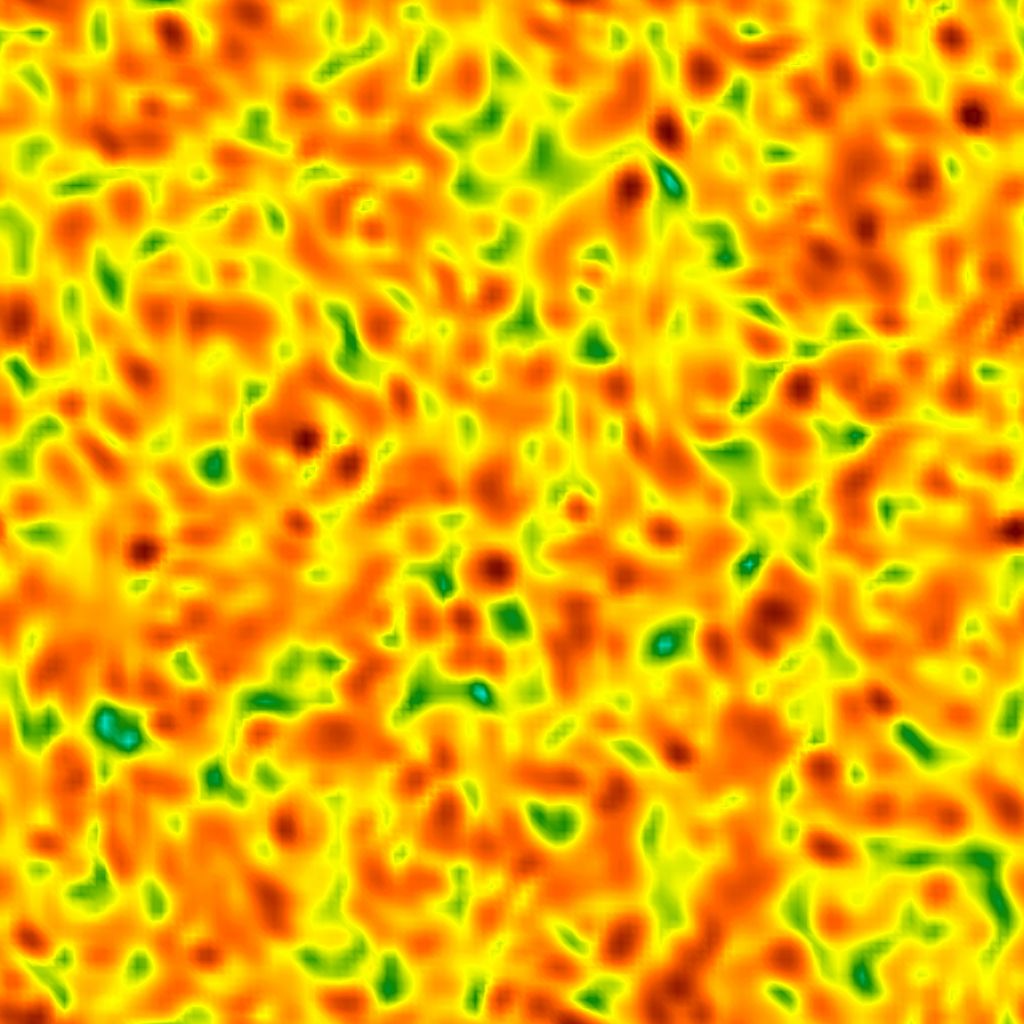}}
\hspace{5pt}
\subfloat{\includegraphics[width=0.2\textwidth]{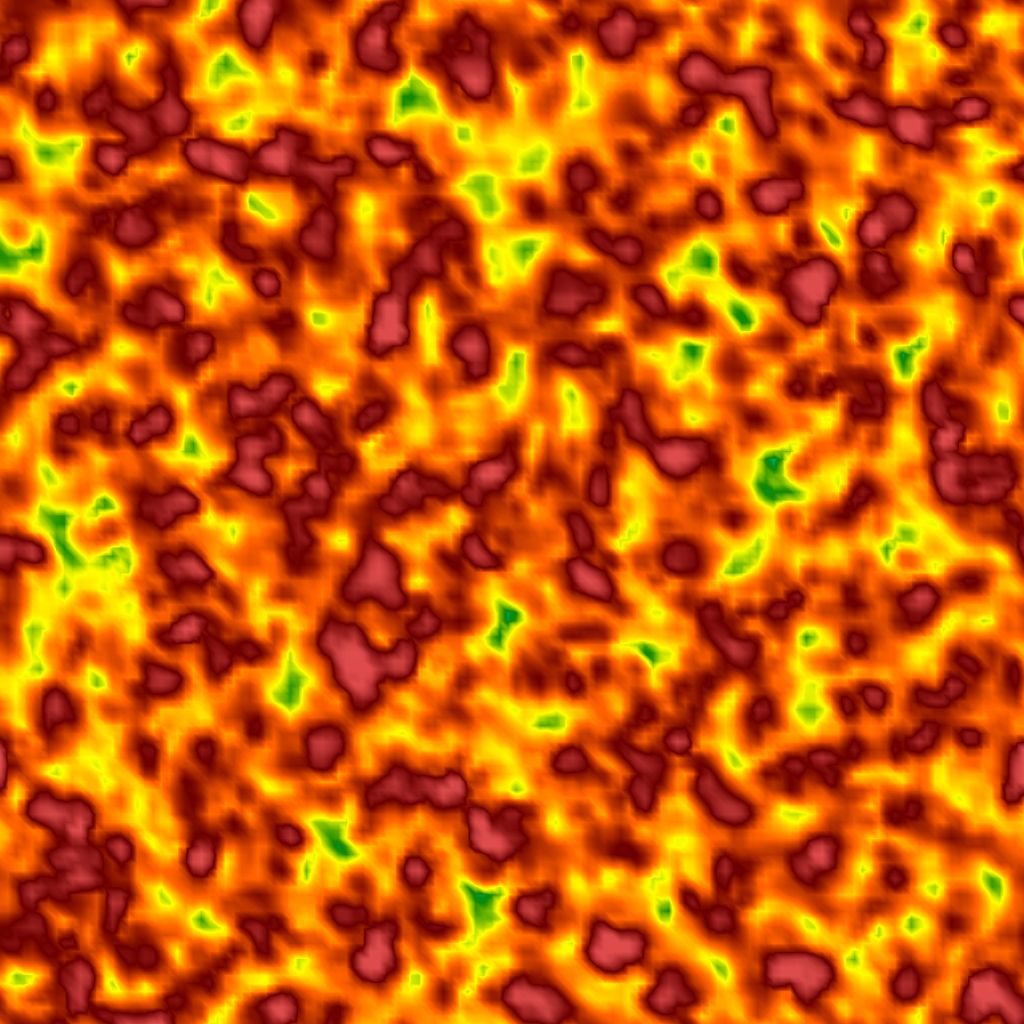}}
\hspace{5pt}
\subfloat{\includegraphics[width=0.2\textwidth]{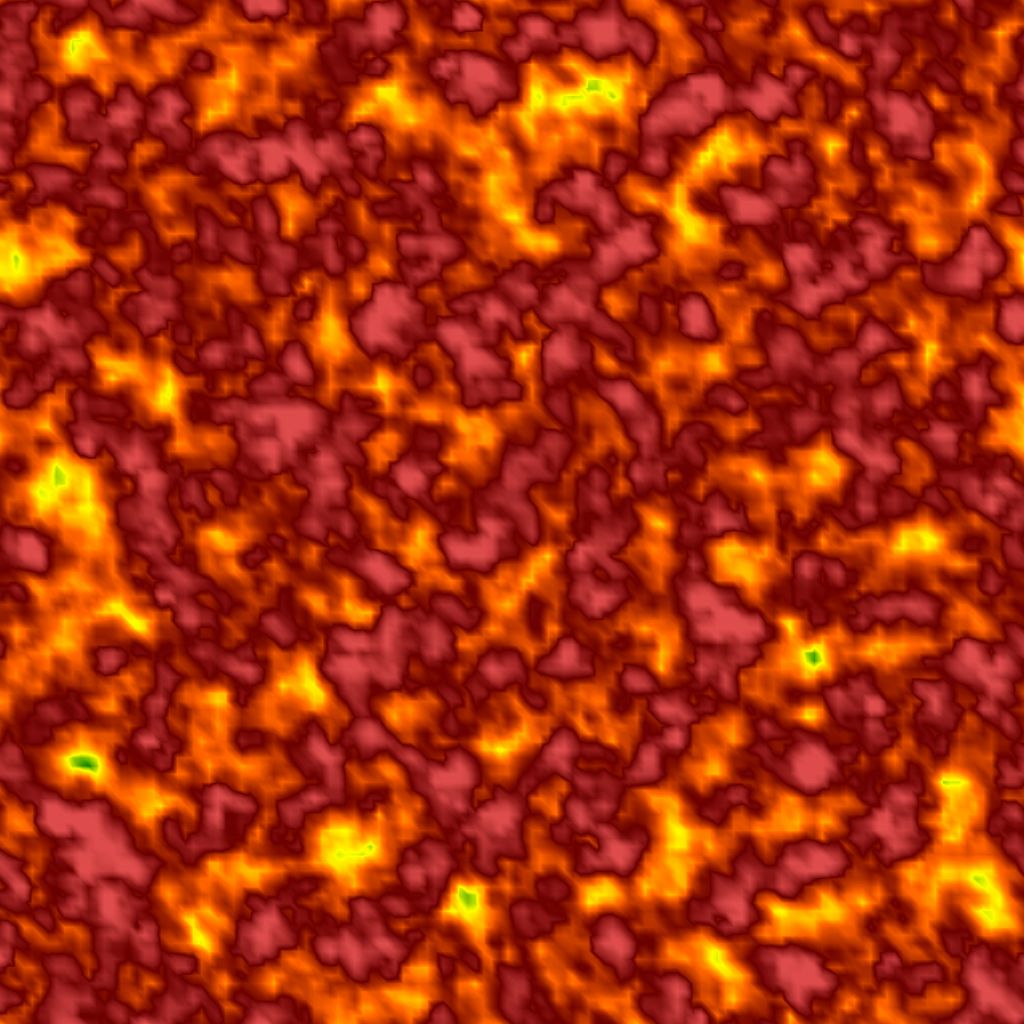}} \\
\smallskip
Abelian \\
\subfloat{\includegraphics[width=0.1\textwidth]{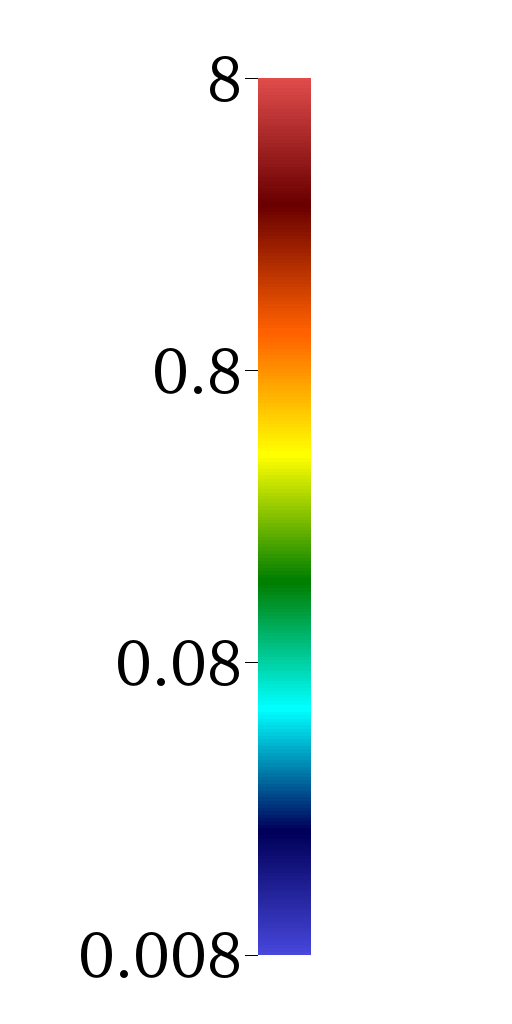}}
\setcounter{subfigure}{0}
\subfloat[$z=8.4$]{\includegraphics[width=0.2\textwidth]{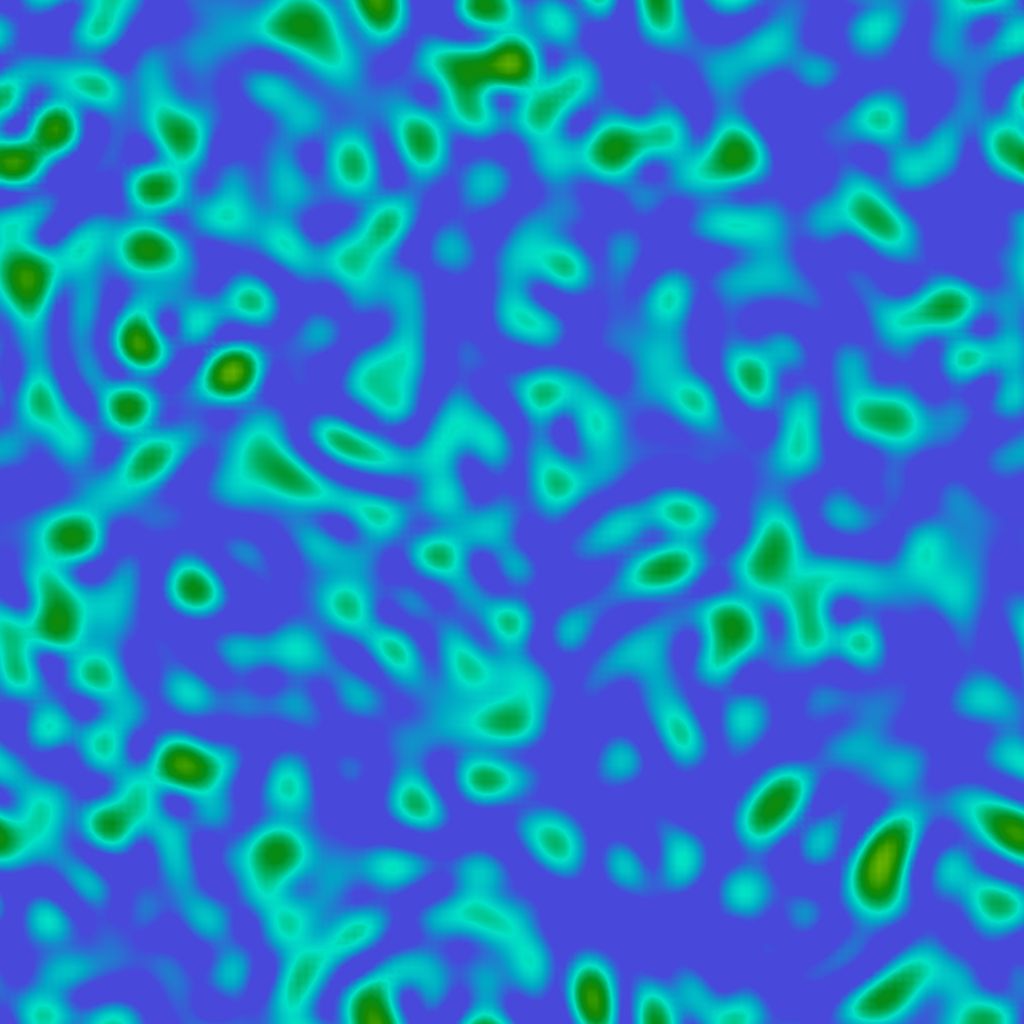}}
\hspace{5pt}
\subfloat[$z=16.4$]{\includegraphics[width=0.2\textwidth]{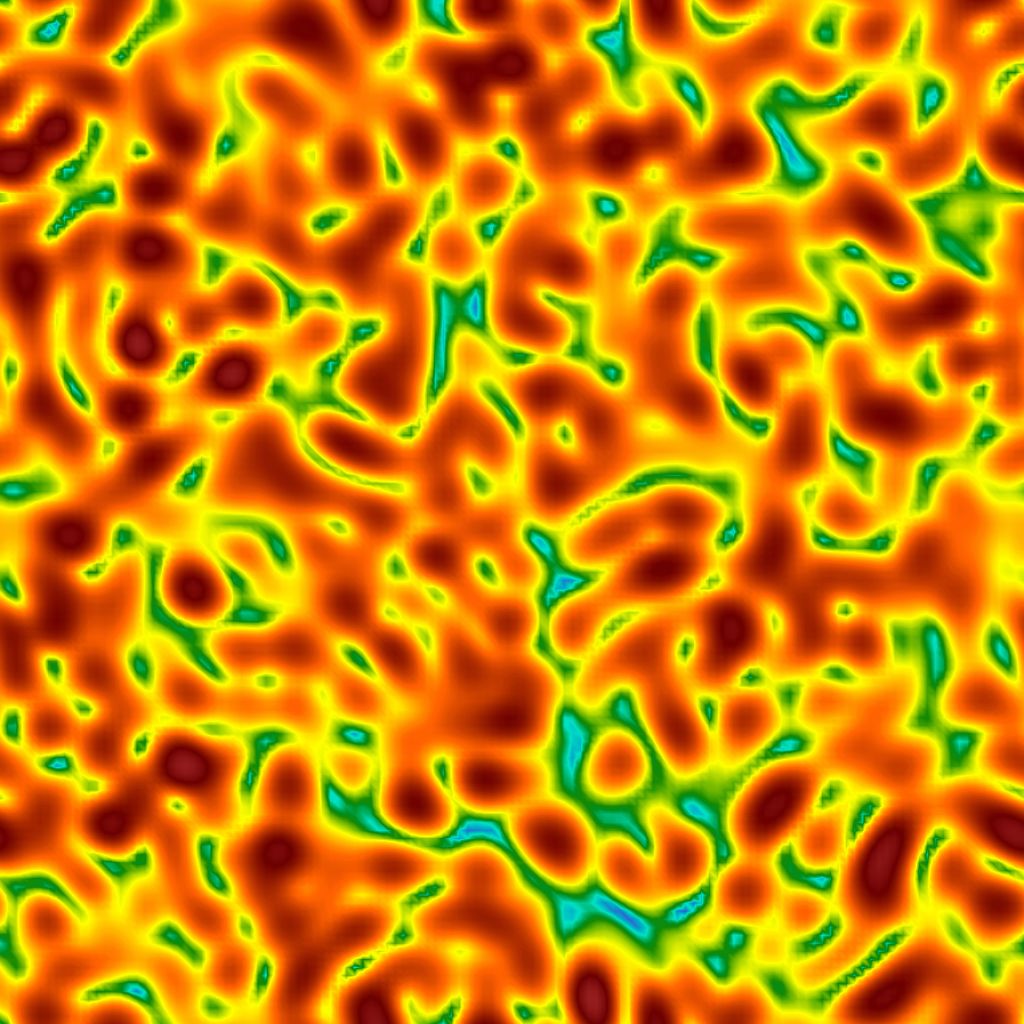}}
\hspace{5pt}
\subfloat[$z=24.4$]{\includegraphics[width=0.2\textwidth]{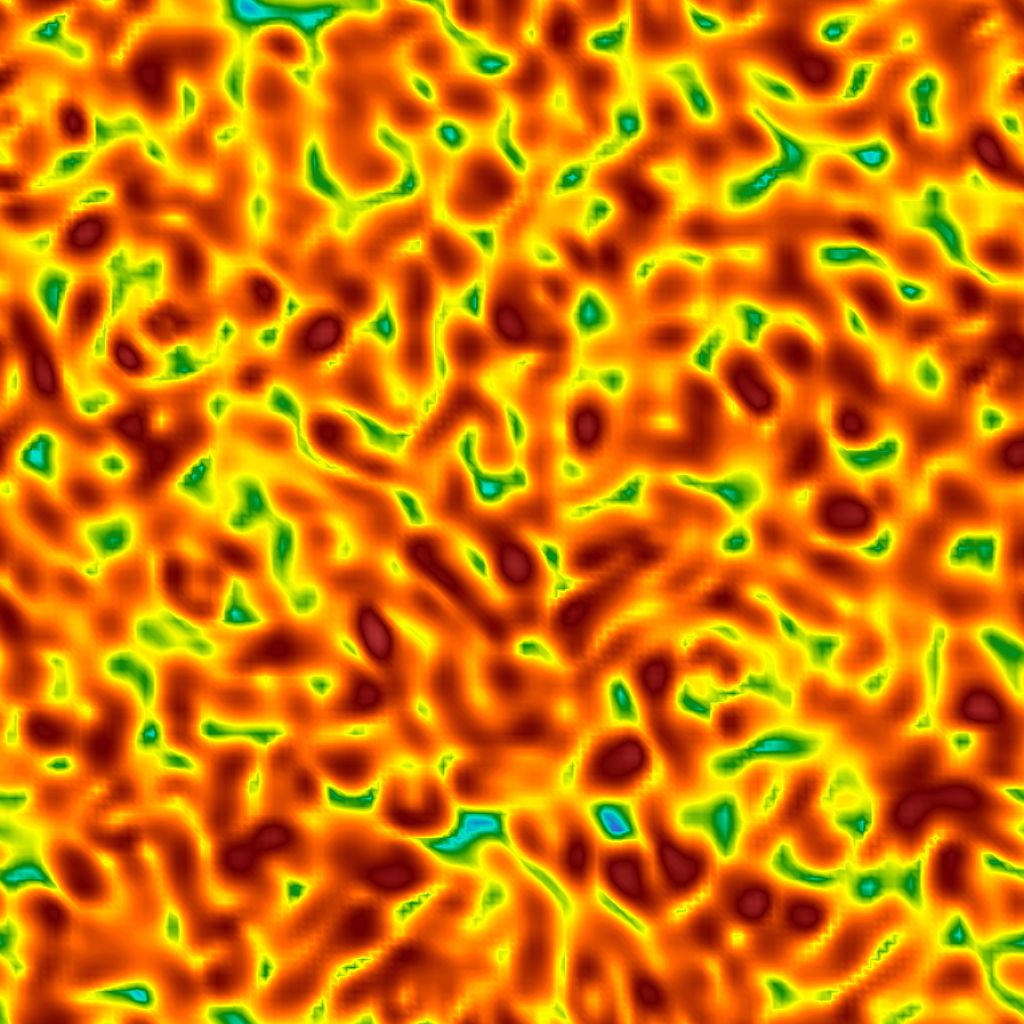}}
\hspace{5pt}
\subfloat[$z=32.4$]{\includegraphics[width=0.2\textwidth]{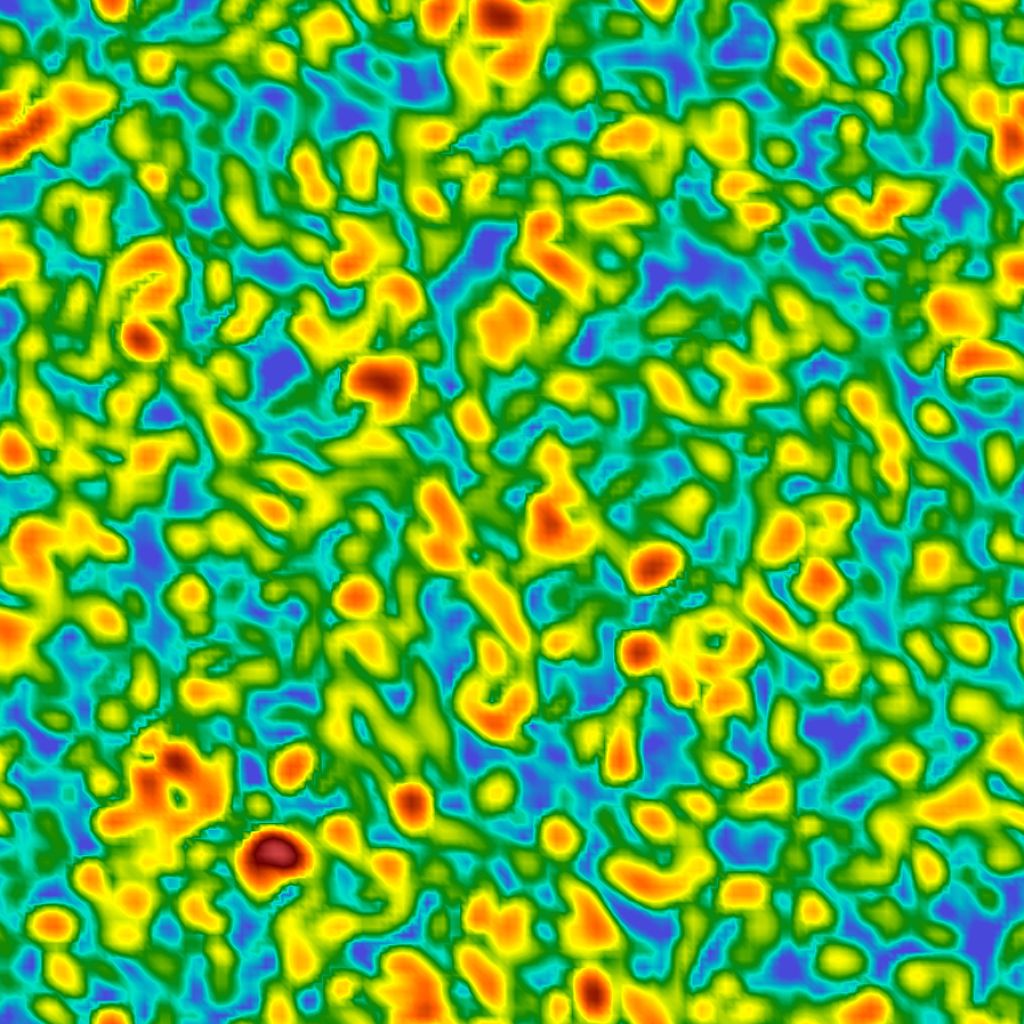}} \\
\caption{\label{fig:gauge_energies}\small
 Slices of the gauge field
  energy density, comparison between Abelian and non-Abelian
  simulations, $q=6.12$, at successive time-slices. Backreaction for the non-Abelian case has
  clearly started before $z \approx 24$ and the non-Abelian self-interaction means
  that by $z\approx 32$ the field configuration is no longer `smooth'. In
  contrast, the Abelian simulation is still dominated by resonant
  structures at $z \approx 32$. The factor of three difference in the legend
  normalizations is due to the additional non-Abelian colour degrees
  of freedom.}
\end{figure}

Figure~\ref{fig:comparison}
depicts the time-evolution of the non-Abelian gauge field particle
numbers. For the sake of comparison, we have also shown the time
evolution in an Abelian case with equivalent Higgs and gauge coupling
strengths. The initial evolution is similar in both cases but a
drastic difference is observed as the particle numbers become much
larger than unity. After this point the non-Abelian distribution is rapidly
extended towards higher momenta through mutual interactions of the
gauge bosons. In the Abelian case, where the interactions are absent
the distribution on the other hand remains strongly peaked around the
momenta $\kappa\lesssim q^{1/4}$ excited by the resonance. Similar
behaviour can also be observed in the gauge field energy densities
illustrated both for the Abelian and non-Abelian case in
Figure~\ref{fig:gauge_energies}

\begin{figure}
\centering
\vspace{-15pt}
\hspace{-50pt}\includegraphics[width=0.67\textwidth]{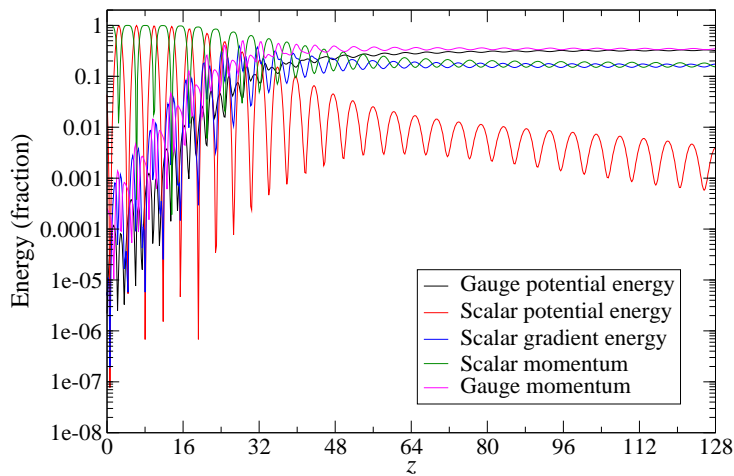}
\caption{\label{fig:energies}\small Time evolution of the individual energy components for a
  typical non-Abelian simulation ($q\approx 6.12$).}
\end{figure}

The bath of non-Abelian gauge particles with momenta extending to
$\kappa\gg q^{1/4}$ rapidly destroys the Higgs condensate through
scattering processes. This can be observed in
Figure~\ref{fig:energies}
which depicts the evolution of the various energy components of the
Higgs and $\mathrm{SU}(2)$ gauge fields. Comparing
Figures~\ref{fig:comparison} and \ref{fig:energies} we clearly see
that the Higgs condensate rapidly starts to decay after the gauge
boson particle numbers at the resonance peak have ceased their
exponential growth. The resonance shuts off and the remnants of the
Higgs condensate decay completely within a few oscillation cycles.

\begin{figure}
\centering
\vspace{-30pt}
\hspace{-50pt}\includegraphics[width=0.75\textwidth]{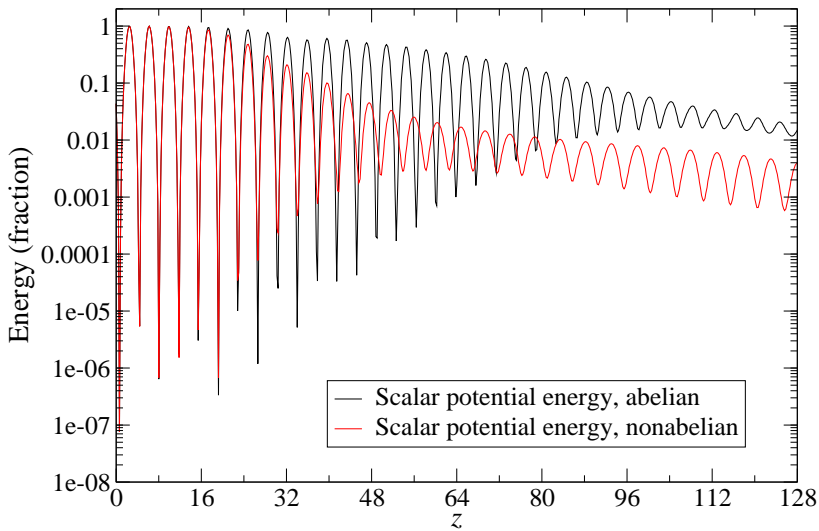}
\caption{\label{fig:potcomp}\small Comparison of scalar
  potential energies for Abelian and non-Abelian simulations ($q\approx 6.12$). The decay of the Higgs condensate is substantially faster in the non-Abelian case.}
\end{figure}

In the Abelian case, a significant fraction ${\cal O}(1)$ of the
condensate remains even after the end of resonance. This is
illustrated in Figure~\ref{fig:potcomp} where we compare the evolution of
the condensate energy in the Abelian and non-Abelian case.
As discussed above, in the Abelian case the gauge field distribution
remains peaked around the long-wavelength modes $k\lesssim q^{1/4}$
excited by the resonance. Correspondingly, the gauge boson
scatterings of the Higgs field are not sufficient to destroy the
condensate and the Higgs remains close to its initial non-equilibrium configuration.

\begin{figure}
\vspace{-25pt}
\subfloat{
\includegraphics[width=0.45\textwidth]{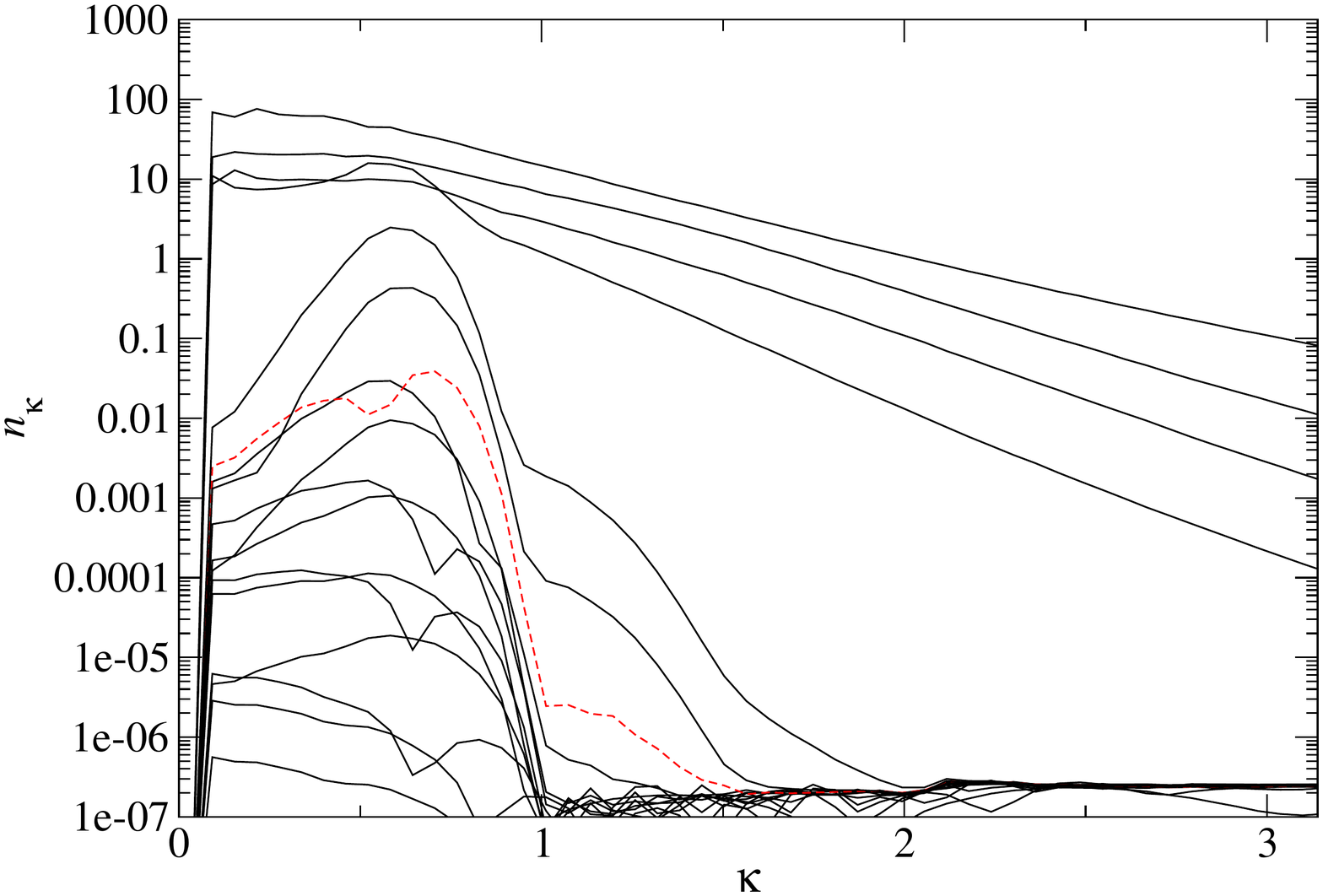}}
\quad
\subfloat{
\includegraphics[width=0.45\textwidth]{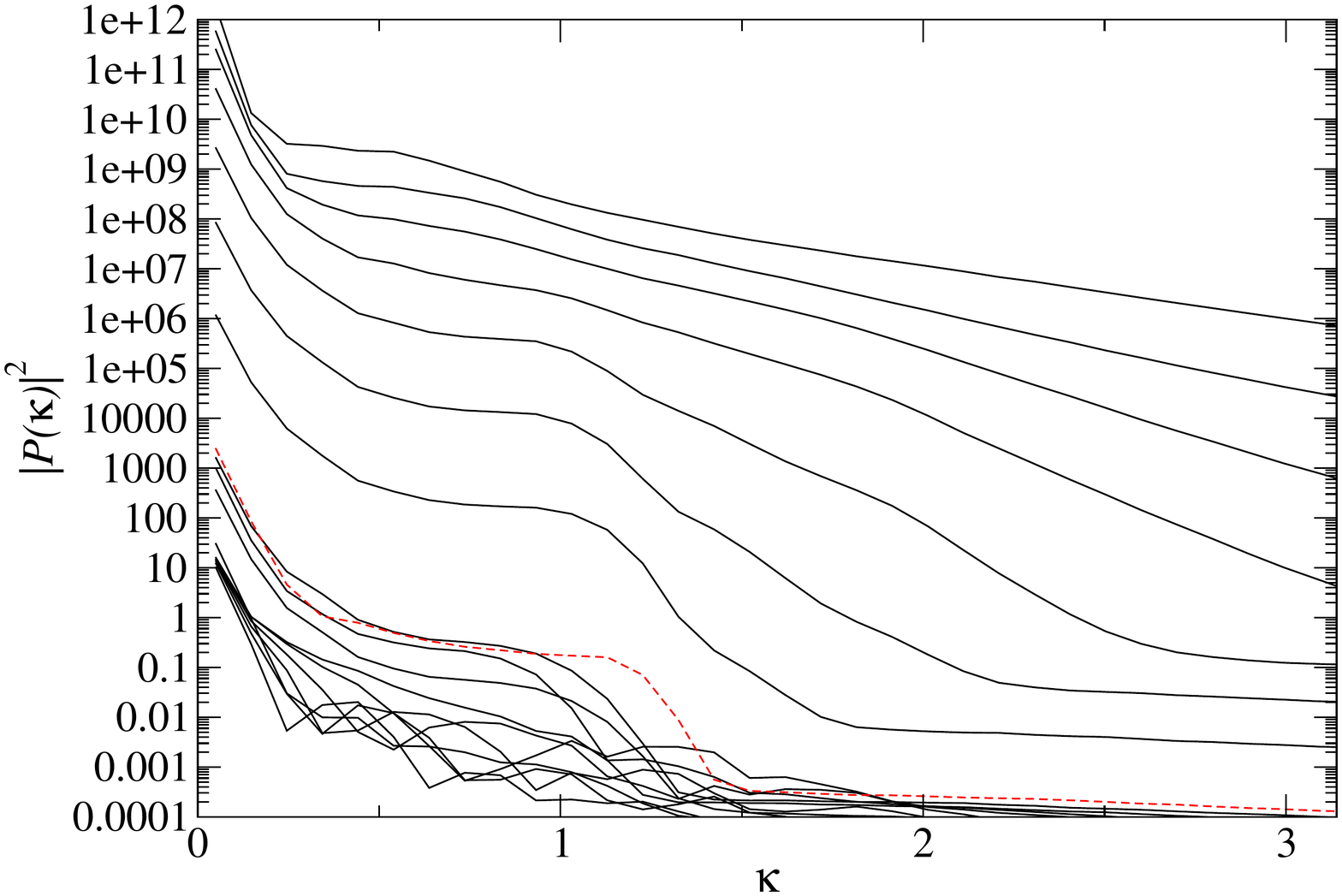}}
\caption{\label{fig:spectrum}\small Plots showing (at left) the particle numbers $n_\kappa$ and
  (at right) the power spectrum of the non-Abelian gauge field, at
  equal times (intervals of $\delta z = 4$) throughout a typical
  simulation ($q\approx 6.12$). The power spectrum closest to $z=0$,
  at $z=0.4$ is shown in both plots as a red dashed line.}
\end{figure}

Finally, in Figure \ref{fig:spectrum} we show the development of particle number for a non-Abelian
simulation at finer time intervals than is possible in figure 2, as well as the gauge field power
spectrum for the same intervals, to give a gauge-invariant picture of the energy deposited in
the bosons.

Note that the results presented here (and in particular
Figure~\ref{fig:comparison}) bear a strong resemblance to Figure 7
in Ref.~\cite{Bodeker:2007fw}. The underlying physics of
equilibration and, presumably, subsequent thermalisation is very
similar. Corresponding results have also been  obtained for tachyonic
preheating in Ref.~\cite{Skullerud:2003ki}.

\section{Conclusions}
\label{sec:conclusions}

In this work we have studied the non-perturbative decay of the primordial Higgs
condensate after inflation into $\mathrm{SU}(2)$ gauge bosons which are
produced through parametric resonance once the condensate leaves
slow-roll and starts to oscillate. We performed a full lattice
simulation of the resonant production of gauge bosons focusing in
particular on the detailed dynamics after the onset of backreaction
and carefully accounting for the effects arising from non-Abelian
interactions.

Resonant decay of the condensate results in the production of
particles within particular bands of momenta in the infrared region.
However, we have found that the scattering and decay processes
resulting from non-Abelian interactions rapidly extend the momentum
distribution into the ultraviolet, which efficiently destroys the
remnants of the Higgs condensate after the onset of backreaction.
This is in sharp contrast to the purely Abelian case where the
distribution remains in the infrared for much longer and the
produced particles do not efficiently destroy the condensate so that
a sizeable part of it survives after the resonance is terminated.

The time for backreaction to set in and the Higgs condensate to
decay is found to be logarithmically dependent on the inflationary
scale $H_*$ which determines the Higgs initial conditions. For
$H_{*}=10^{8}$ GeV we find that $90$\% of the Higgs condensate has
decayed after $n\sim 10$ oscillation cycles.

After the decay is completed the system starts to evolve towards  
equilibrium. Energy is rapidly transferred from the initially occupied  
IR modes towards higher momenta and the system appears to evolve  
towards a stationary state. This resembles the self-similar behaviour  
observed in earlier lattice studies of preheating \cite{Micha:2002ey,Micha:2004bv} and also studies of  
heavy ion collisions, see e.g. \cite{Berges:2013eia, Kurkela:2011ti}. The  
evolution of this stage towards the full thermal equilibrium can  
however not be analysed within a classical simulation which fails in  
describing the UV modes. We leave a more detailed investigation of the  
thermalisation for a future work.

We have emphasized that unless the SM Higgs sector is significantly
modified by new physics, a Higgs condensate is inevitably generated
during inflation. Thus it is important to understand the details of
its decay as it may be significant for subsequent physics. The Higgs
condensate sets specific out-of-equilibrium initial conditions for
the hot Big Bang epoch which could have significant ramifications
ranging from primordial perturbations and baryogenesis to
non-thermal production of dark matter \cite{DeSimone:2012qr,
DeSimone:2012gq,Choi:2012cp,Enqvist:2014zqa, Kusenko:2014lra}. In
all these cases the decay time of the Higgs condensate is a crucial
factor affecting the observational impacts. Our results indicate
that, in the absence of thermal bath, once the Higgs condensate
unfreezes and starts to oscillate it decays completely after ${\cal
O}(10)$ oscillation cycles.

While we have considered the decay of the condensate within the context
of the Standard Model it would be interesting to extend the analysis
to the case of Higgs inflation, where the condensate plays the role of the
inflaton through a large non-minimal coupling to gravity. While
order of magnitude corrections to previous estimates \cite{Bezrukov:2008ut,GarciaBellido:2008ab} are
not expected, a proper inclusion of non-Abelian interactions
appears crucial to precisely determine the predicted reheating
temperature of the scenario. Other extensions of the  Standard Model
may also be of interest. For instance, in some extensions the resonance might turn out to be narrow rather
than broad. In such a case the non-Abelian nature of the interactions
is expected to become important already before the onset of
backreaction \cite{Enqvist:2014tta}.

\acknowledgments

The numerical simulations were performed on the Norwegian computer
cluster Abel, under the NOTUR project. SN is supported by the Academy
of Finland grant 257532. SR is supported by the Magnus Ehrnrooth
Foundation.  DJW is supported by the People Programme (Marie
Sk{\l}odowska-Curie actions) of the European Union Seventh Framework
Programme (FP7/2007-2013) under grant agreement number
PIEF-GA-2013-629425. We acknowledge useful discussions with Tommi
Markkanen, Kari Rummukainen and Anders Tranberg.

\appendix
\section{Details of lattice implementation}
\label{sec:latticedetails}

The lattice Hamiltonian is
\begin{multline}
H = \sum_x \pi^\dag(x) \pi(x) + \sum_{x,i,a} \frac{1}{2} P_i^a(x)
P_i^a(x) + \frac{2}{g^2} \sum_{x,i<j} \left[ 2-\mathrm{Re} \, \mathrm{Tr}\, U_i(x) U_j(x+\hat\imath) U_i^\dag (x+\hat\jmath) U_j^\dag (x) \right] \\
+ \sum_{x,i} \left[ 2\phi^\dag(x) \phi(x) - 2\mathrm{Re}\,\phi^\dag(x) U_i(x) \phi(x+\hat\imath) \right] + \sum_x V(\phi^\dag(x) \phi(x))
\end{multline}
where $\pi$ and $P$ are the conjugate momenta for the Higgs and gauge
fields, $U_i(x)$ are the link variables corresponding to the gauge
field (see below) and $\phi$ is the Higgs field.
The equations of motion are then
\begin{align}
\phi(t + \delta t, x) & = \phi(t, x) + \delta t \, \pi(t + \delta t/2, x) \\
U_i(t+\delta t, x) & = \exp\left[ - i \frac{g}{2} P_i^a(x,t + \delta t/2) \sigma^a \, \delta t\right] U_i(t,x) \\
\pi(t+\delta t/2, x) & = \pi(t-\delta t/2,x) \nonumber \\
& \qquad + \delta t \left[ \sum_i \left[ U_i(t,x)\phi(t,x+\hat\imath)
    - 2\phi(t,x) + U_i^\dag(t,x-\hat\imath)\phi(t,x-\hat\imath)\right]
  - \frac{\partial V}{\partial \phi^\dag} \right] \\
P_k^m(t+\delta t/2, x)  & = P_k^m(t-\delta t/2, x) + \delta t\left[ g \, \mathrm{Re} \left[ \phi^\dag(t,x+\hat k) U^\dag_k(t,x) i \sigma^m \phi(t,x)\right] \vphantom{\frac{1}{g}} \right. \nonumber \\
& \qquad  \left. - \frac{1}{g} \sum_i \mathrm{Tr} \left[ i \sigma^m U_k(t,x)U_i(t,x+\hat k) U_k^\dag(t,x+\hat\imath) U_i^\dag (t,x) \nonumber \right. \right. \\
& \qquad \qquad \left. \left. + i\sigma^m U_k(t,x) U_i^\dag(t,x+\hat k -\hat \imath) U_k^\dag(t,x-\hat\imath) U_i(t,x-\hat\imath) \right] \vphantom{\frac{2}{g}} \right]
\end{align}
on the lattice, with Gauss law
\begin{multline}
G(x) = \sum_i \mathrm{Tr} \left[ i \,  \sigma^k P_i(t +
  \delta t/2, x) - i \, \sigma^k U_i^\dag(t, x-\hat\imath) P_i(t + \delta t/2, x-\hat \imath) U_i(t,x-\hat\imath) \right] \\
 + 2 g \, \mathrm{Re} \left[ \pi^\dag (t + \delta t/2, x) i \sigma^k
   \phi(t,x) \right] = 0.
\end{multline}
See for example Refs.~\cite{Rajantie:2000nj,DiazGil:2008tf,Skullerud:2003ki} for
more details.

Our measurements of non-Abelian gauge boson particle number use a lattice
approximation to $A_i^a(x,t)$. Given the parallel transporter is
defined through
\begin{equation}
U_i(x,t) = \exp\left[-i g \frac{\sigma^a}{2} A_i^a (x+\hat\imath/2,t) \right]
\end{equation}
we deduce
\begin{equation}
\label{eq:approxcontfield}
A_i^a(x,t) = \frac{i}{g} \mathrm{Tr} \,  \sigma^a U_i(x,t)
\end{equation}
to second order in the lattice spacing.

\subsection{Gauge fixing in \texorpdfstring{$\mathrm{SU}(2)$}{SU(2)}}
\label{sec:gaufix}

Our equations of motion are in temporal gauge, and so while at some
point the system may be in Coulomb gauge, the time evolution of the
system will generally take it away from this form. We fix Coulomb
gauge on the lattice~\cite{Fachin:1991pu,Giusti:2001xf}, when we wish
to measure particle numbers. To do this we must find a lattice gauge
transformation $g(x) \in \mathrm{SU}(2)$ such that
\begin{equation}
F^g[U] = \frac{1}{6 V} \sum_{x,i} \mathrm{Re} \, \mathrm{Tr} \, U^g_i(x)
\end{equation}
is stationary, where
\begin{equation}
U^g_i(x) \equiv g(x) U_i(x) g^\dag(x +\hat\imath).
\end{equation}
We look for a maximum of this expression by carrying out a
combination of simulated annealing and overrelaxation steps, a
strategy that is widely used for lattice quantum field theory
simulations where fixing to Coulomb, Landau or Abelian gauge is
required (see for example
Refs.~\cite{Giusti:2001xf,Bali:1994jg,Schrock:2012fj}). We perform
10000 sweeps of simulated annealing combined with overrelaxation. The
divergence $\Delta(x)$ of the equivalent of the continuum gauge field,
Eq.~(\ref{eq:approxcontfield}), computed through
\begin{equation}
\Delta(x) = \sum_i \left(A_i(x) - A_i(x-\hat{\imath})\right)
\end{equation}
can be normed to give a measure $\theta$ of the quality of the gauge
fixing,
\begin{equation}
\theta = \frac{1}{3V} \sum_x \mathrm{Tr} \, \Delta(x) \Delta^\dag(x).
\end{equation}
After our gauge fixing procedure we typically obtain $\theta \approx
10^{-11}$ given an initial value around $10^{-2}$.

The above algorithm as implemented in our code is fully parallelised
so the need to fix the gauge to produce meaningful particle number
results is, overall, not a significant burden.

To obtain the electric field
$E_i^a(x,t)$ note that the Coulomb gauge condition is
time-independent. We therefore carry out the gauge fixing procedure --
with the same $g(x)$ -- on
plaquettes at two sequential timesteps $U^g_i(x,t)$ and
$U^g_i(x,t+\delta t)$ and use these to approximate the electric field
$E_i(x,t)$ through
\begin{equation}
\exp \left[ - i\frac{g}{2} E_i^a(x,t) \sigma^a \, \delta t\right]
\approx U^g_i(x,t+\delta
t) U^{g\dag}_i (x,t).
\end{equation}
Note that the gauge transformation $g(x)$, while time independent,
does not necessarily bring configurations on any other timestep into
Coulomb gauge, so our estimation of the gauge-fixed electric field
could have $\mathrm{O}(\delta t)$ errors.

Having obtained lattice approximations to $E_i^a(x,t)$ and
$A_i^a(x,t)$ we can compute Eqs. (\ref{eq:Aexpectation}-\ref{eq:Eexpectation}) and obtain
particle numbers for the system at a given time from Eq.~(\ref{eq:particlenumber}).

Note that, as demonstrated previously in Ref.~\cite{Skullerud:2003ki},
at large particle numbers the gauge fixing procedure does not
substantially affect the final results.  One could infer the same
conclusions as we reach in this paper from the transverse `particle
number' obtained without gauge fixing; the discrepancy is around
$10\%$.

\subsection{Gauge fixing in \texorpdfstring{$\mathrm{U}(1)$}{U(1)}}

The same principle is applied here, except that a simple gradient flow
procedure easily determines the transformation $A_i(x,t) \to A_i'(x,t)
+ \Lambda(x,t)$ that satisfies the condition $\partial_i A_i = 0$. The
same gauge transformation $\Lambda$ can then be applied to the
following timestep to obtain $E_i(x,t)$ in a manner analogous to that
given above.

\subsubsection{Numerical tests}

We confirmed that our backreaction times and other principal results
were logarithmic in the initial energy density $\xi$. We also checked
that the results did not depend on the timestep size or the lattice
volume $V$ (we tested $128^3$ and $256^3$, both of which capture the
parametric resonance well). While we can test the large-volume and
infrared robustness of our results, the system is conformal and there
is no mass scale. Therefore we cannot take the continuum limit or draw
any conclusions about physics once particle numbers in the UV start to
grow after backreaction.

The simulations should always satisfy the Gauss law. The worst-case
total Gauss law violation $\sum_x G(x)^2 / V$ in the non-Abelian case was
around $10^{-10}$, and occurred around the time of backreaction.

\bibliographystyle{jhep.bst}
\bibliography{HiggsBib.bib}

\end{document}